\documentclass{easychair}
\usepackage{rotating}
\usepackage{pdflscape}
\usepackage{marvosym}
\usepackage{multirow}

\usepackage{listings}
\begin{document}

%% Front Matter
%%
% Regular title as in the article class.
%
\title{Attacks Against Security Context in 5G Network}

% \titlerunning{} has to be set to either the main title or its shorter
% version for the running heads. Use {\sf} for highliting your system
% name, application, or a tool.
%
\titlerunning{Attacks Against Security Context}

\volumeinfo
	{The 6th International Symposium on Mobile Internet Security (MobiSec'22), December 15-17, 2022, Jeju Island, Republic of Korea}
	{Article No. 1}        

% Authors are joined by \and and their affiliations are on the
% subsequent lines separated by \\ just like the article class
% allows.
%
\author{\\
Zhiwei Cui${}^{1}$, Baojiang Cui${}^{1}$\thanks{Corresponding author: School of Cyberspace Security, Beijing University of Posts and Telecommunications, Beijing, China, Email: cuibj@bupt.edu.cn}~, Li Su${}^{2}$, Haitao Du${}^{2}$, Hongxin Wang${}^{1}$, and Junsong Fu${}^{1}$\\
${}^{1}$Beijing University of Posts and Telecommunications, Beijing, China\\
zwcui@bupt.edu.cn, cuibj@bupt.edu.cn, wanghongxin@bupt.edu.cn, fujs@bupt.edu.cn\\
${}^{2}$China Mobile Research Institute, Beijing, China\\
suli@chinamobile.com, duhaitao@chinamobile.com
}

% \authorrunning{} has to be set for the shorter version of the authors' names;
% otherwise a warning will be rendered in the running heads.
%
\authorrunning{Z. Cui et al.}

\maketitle

%------------------------------------------------------------------------------
% Abstract
%
\begin{abstract}
\noindent
The security context used in 5G authentication is generated during the Authentication and Key Agreement (AKA) procedure and stored in both the user equipment (UE) and the network sides for the subsequent fast registration procedure. Given its importance, it is imperative to formally analyze the security mechanism of the security context. The security context in the UE can be stored in the Universal Subscriber Identity Module (USIM) card or in the baseband chip. In this work, we present a comprehensive and formal verification of the fast registration procedure based on the security context under the two scenarios in ProVerif. Our analysis identifies two vulnerabilities, including one that has not been reported before. Specifically, the security context stored in the USIM card can be read illegally, and the validity checking mechanism of the security context in the baseband chip can be bypassed. Moreover, these vulnerabilities also apply to 4G networks. As a consequence, an attacker can exploit these vulnerabilities to register to the network with the victim's identity and then launch other attacks, including one-tap authentication bypass leading to privacy disclosure, location spoofing, etc. To ensure that these attacks are indeed realizable in practice, we have responsibly confirmed them through experimentation in three operators. Our analysis reveals that these vulnerabilities stem from design flaws of the standard and unsafe practices by operators. We finally propose several potential countermeasures to prevent these attacks. We have reported our findings to the GSMA and received a coordinated vulnerability disclosure (CVD) number CVD-2022-0057.\\
\newline
\newline
\textbf{Keywords}: Security Context, 5G Network, Attack 
\end{abstract}
%------------------------------------------------------------------------------
\section{Introduction}
The Universal Subscriber Identity Module (USIM) card stores critical authentication credentials (e.g., identities and permanent authentication keys) for subscribers to access the 5G network to consume services, such as the Internet, voice, short message service (SMS), etc. The Authentication and Key Agreement (AKA) procedure utilizes these credentials to complete key derivation and mutual authentication between the user equipment (UE) and the network to ensure communication security. Usually, a security context is created as the result of the AKA procedure and maintained by the UE and the network. Then the UE can use this security context to quickly register to the network in a following period without going through the AKA procedure. Given its importance, it is hence necessary to investigate the existing design and implementation of the security context in the 5G network.

The UE consists of the USIM card and the mobile equipment (ME). The 3rd Generation Partnership Project (3GPP) stipulates that the security context can be stored in the USIM or ME \cite{3gpp.24.501, 3gpp.24.301}. If the security context is not present on the USIM, it is stored in the baseband chip in the ME together with the permanent identity. Moreover, 3GPP has proposed some protection measures to defend against attacks on the security context. The PIN-based access control can prevent malicious access to the security context in the USIM card \cite{3gpp.31.102}. The security context stored in the baseband chip can only be used if the permanent identity from the USIM matches the identity stored in the non-volatile memory. And the standard specifies that the security context should be deleted from the baseband chip in some scenarios, such as the USIM is removed from the ME when the ME is in power on state \cite{3gpp.33.401, 3gpp.33.501}. However, whether these methods can achieve the intended purpose and whether there will be shortcomings in the implementation stage are still worth discussing.

There is already some work involving the analysis of the security context. Some basic concepts of the security context are discussed in \cite{schneider2015towards}. Through a man-in-the-middle (MITM) attack, the victim continuously deletes the current security context and initiates the initial registration request, which will accelerate battery consumption \cite{shaik2019new}. The security context stored in the USIM card can be extracted by hardware or software to initiate several attacks \cite{zhao2021securesim}, which proves that the PIN-based access control is flawed. However, the related works are still of some limitations: (A) Some studies did not pay attention to the security of the security context itself \cite{schneider2015towards, shaik2019new}; (B) Another situation that the security context stored in the baseband chip is not considered \cite{zhao2021securesim}; (C) These studies were conducted through ad-hoc manual analyses \cite{schneider2015towards, shaik2019new, zhao2021securesim}.

In this paper, we systematically model the fast registration procedure based on the security context using the formal method. And we use ProVerif, an automatic symbolic protocol verifier \cite{blanchet2016modeling}, to automatically investigate whether the safeguards proposed by 3GPP for the security contexts under multiple storage methods can achieve their intended purpose. Unfortunately, the results yield a negative answer. We have uncovered two vulnerabilities, one of which is novel. It is worth mentioning that we have proposed a more efficient way to exploit the already discussed vulnerability \cite{zhao2021securesim}. An attacker could exploit these vulnerabilities to register to the network as the victim. Additionally, we discuss further attacks (e.g., forging location information and stealing privacy) that an attacker can launch after registering to the network. Our analysis reveals that these vulnerabilities are caused by design weaknesses of the 3GPP standard and unsafe practices by operators. In the end, we discuss potential countermeasures.

The contributions of this paper are summarized as follows:
\begin{itemize}
\item We present a comprehensive formal model for the fast registration procedure using the security context. And we identify two vulnerabilities, with one that has not been reported before. We have disclosed the novel vulnerability to the GSMA, resulting in 1 vulnerability ID (CVD-2022-0057).
\item We devise the impersonation attack against mobile users by exploiting the identified vulnerabilities. And we have responsibly confirmed our attacks through experimentation in three operators. Moreover, we assess the impact of the attack, such as one-tap authentication bypass leading to privacy disclosure, location spoofing, etc.
\item We analyze the root causes of the vulnerabilities and propose several potential countermeasures to prevent these attacks from multiple perspectives.
\end{itemize}

The rest of this paper is organized as follows: We present the preliminaries of USIM, 5G registration procedure, and 5G security context that are relevant to our research in Section~\ref{sect:background}. Section~\ref{sect:modeling} introduces our interpretation and formal model of the fast registration procedure based on the security context in ProVerif. We present the vulnerabilities discovered through our model in Section~\ref{sect:vulnerability}. Then we design several attacks as described in Section~\ref{sect:attack}. Section~\ref{sect:evaluation} describes a survey to evaluate the vulnerabilities and attacks. Several countermeasures are presented in Section~\ref{sect:countermeasure}. We summarize the related work in Section~\ref{sect:relatework}. At last, Section~\ref{sect:conclusion} concludes this paper.
%------------------------------------------------------------------------------
\section{Background}
\label{sect:background}
\subsection{USIM}
The USIM card is a smart card specified in \cite{3gpp.31.102} and contains a microprocessor, three types of memory, which are RAM, ROM and EEPROM, and some security functions. Each USIM card has a unique integrate circuit card identity (ICCID). The USIM card can be viewed as a black box interfacing with ME through the standard application programming interface (API) \cite{savoldi2007sim}. A mobile user can access the 5G network through the ME equipped with a valid USIM card to consume services, including the Internet, voice, and SMS.

The USIM card stores various critical information for a subscriber, such as identities, permanent authentication keys, security contexts, phone number, etc. These parameters are stored in the form of files. Each file has a unique identifier and can be set with different access conditions determined by the mobile carrier. The access conditions indicated in the file header can be divided into 4 types \cite{zhao2021securesim, savoldi2007sim}: (1) Always (ALW): the command is executable on the file with no restrictions; (2) PIN: the command is executable on the file after entering the correct PIN unless the verification is disabled; (3) ADM: the command is executable on the file after entering the correct administrative key owned by the mobile carrier; (4) Never (NEV): the command cannot executable on the file. The applications (e.g., the baseband chip, card reader, and APP) can send application protocol data unit (APDU) to the USIM card to read or update these files.

In our experiments, a total number of three files will be involved: the international mobile subscriber identity (IMSI) in file IMSI (ID: 6F07), the globally unique temporary identifier (GUTI) in file EPSLOCI (ID: 6FE3), and the 4G Non-Access-Stratum (NAS) security context in file EPSNSC (ID: 6FE4). The 3GPP standard stipulates that the read permission of these files should be set to PIN mode \cite{3gpp.31.102}. If the PIN verification is enabled, the card holder needs to input the correct PIN to unlock the USIM card after inserting the card to a new device or restarting the phone. The length of the PIN is usually set to 4-8 digits and the number of consecutive incorrect PIN entries is limited. If the limit is exceeded, the USIM card will be automatically locked or destroyed.
\begin{figure}[htb!]
	\begin{centering}
	\includegraphics[width=0.7\textwidth]{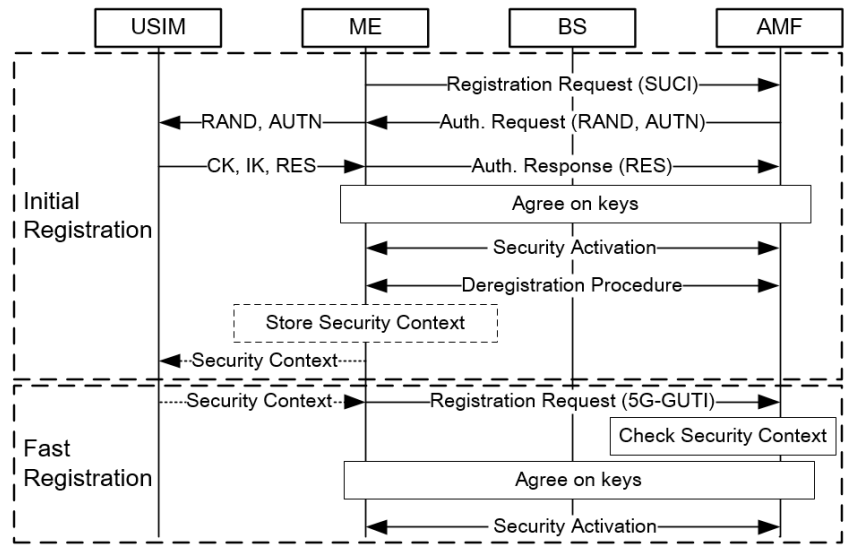}
	\caption{5G registration procedure}
	\label{fig:attach}
	\end{centering}
\end{figure}
\subsection{5G Registration Procedure}
The 5G protocol stacks are divided into the control plane and the user plane. The registration procedure belongs to the control plane, which is the responsibility of the access and mobility management function (AMF). The base station (BS) provides the wireless network for the UE and the AMF to communicate with each other. Figure~\ref{fig:attach} shows the registration procedure \cite{3gpp.24.501, 3gpp.24.301, zhao2021securesim}. The registration procedure includes the initial registration and the fast registration.

\begin{figure}[htb!]
	\begin{centering}
	\includegraphics[width=0.45\textwidth]{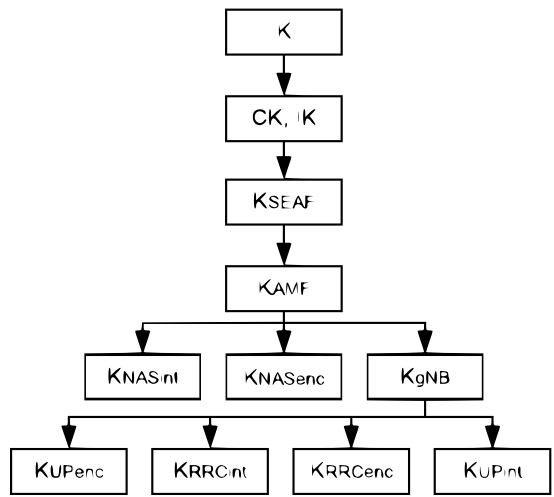}
	\caption{The key derivation}
	\label{fig:key}
	\end{centering}
\end{figure}

\textbf{Initial Registration.} To protect subscriber privacy, the ME initiates an initial registration request message containing the Subscription Concealed Identifier (SUCI) instead of the Subscription Permanent Identifier (SUPI) \cite{3gpp.33.501}. If the AMF confirms that the identity is valid, it initiates the authentication procedure, also known as the AKA procedure. The AMF sends an authentication request message to the ME, containing a random number (RAND) and authentication token (AUTN) generated based on the permanent key and RAND. Then the ME forwards these two parameters to the USIM card in the form of APDU. The USIM card calculates cipher key (CK), integrity key (IK) and authentication response through the permanent key and security algorithms preset in the USIM card, and returns them to the ME. Then the ME and the network derive several keys for data encryption and integrity protection as shown in Figure~\ref{fig:key}. Next, the ME and the network negotiate encryption and integrity protection algorithms to activate security. Finally, when the ME initiates the deregistration procedure (e.g., turns on the airplane mode or powers off), the UE (including the USIM and the ME) and the network maintain a security context simultaneously. For the UE, the security context is stored in the ME or the USIM card.

\textbf{Fast Registration.} The 3GPP standard specifies a fast registration procedure when a 5G security context is stored in the UE \cite{3gpp.24.501, 3gpp.24.301}. The cached context includes 5G-GUTI and 5G NAS security context. If the UE has a cached context, it shall send a registration request message containing 5G-GUTI as an identity, not SUCI. With the 5G NAS security context, the sent message shall be ciphered in a NAS container which is ciphered and shall also be integrity protected \cite{zhao2021securesim}. If the AMF has the same security context, and successfully decrypts the NAS container and verifies the integrity of the registration request message, the AKA procedure may be omitted. Then the UE and the network derive the keys required for subsequent communication based on this security context to activate security. It's worth noting that the 3GPP standard does not mandate skipping the AKA procedure, which depends on the operator. The three operators involved in this paper all provide fast registration services.
\subsection{5G Security Context}
The UE and the network need to establish a security context before activating security. Usually, a security context is generated as the result of an AKA procedure between the UE and the network and maintained by the UE and the network simultaneously. The 5G security context discussed in this paper includes the 5G-GUTI and the 5G NAS security contexts. If the USIM supports storing the 5G security context, then the 5G-GUTI and the 5G NAS security context will be stored in file EF5GS3GPPLOCI (ID: 4F01) and file EF5GS3GPPNSC (ID: 4F03) respectively in the form of files. Otherwise, they will be stored in the baseband chip together with the permanent identity SUPI of the subscriber. The 5G NAS security context contains $K_{AMF}$ with the associated key set identifier (ngKSI), the UE security capabilities, and the uplink and downlink NAS COUNT values. The UE security capabilities identify the selected NAS integrity and encryption algorithms. The UE derives the NAS encryption and integrity keys according to $K_{AMF}$, and uses these algorithms to complete the protection of the registration request message as mentioned above. In addition, the registration request message contains the 5G-GUTI, the ngKSI and the uplink NAS COUNT in plaintext. The AMF queries the corresponding security context through the 5G-GUTI and the ngKSI. The uplink and downlink NAS COUNT values are used to resist replay attacks.

In order to prevent the security context from being used illegally, the 3GPP proposes some security protections \cite{3gpp.31.102, 3gpp.33.401, 3gpp.33.501}: (1) The PIN-based access control prevents malicious reading of the security context stored in the USIM card. (2) In order to use the security context stored in the baseband chip, the permanent identity of the USIM card inserted into the ME shall match the permanent identity corresponding to the security context. It is worth mentioning that the above discussion also applies to the 4G security context. In the following content of this article, if there is no special emphasis on the security context in 4G, we refer to the security context in 5G.
%------------------------------------------------------------------------------
\section{Modeling the Fast Registration Procedure in ProVerif}
\label{sect:modeling}
In this section, we present the formal model for the fast registration procedure. First, we introduce the security assumptions and goals. Then we briefly introduce the formal verification tool ProVerif and present the modeling of security goals and the fast registration procedure.
\subsection{Security Assumptions}
The security assumptions are very strong in the 3GPP specifications \cite{3gpp.33.401, 3gpp.33.501}. However, many deployments of operators do not strictly follow them. To provide a more precise analysis, we take the real-world scenarios into account when defining security assumptions.

\textbf{Assumptions on Cryptographic Algorithms.} We assume that all cryptographic algorithms are public and secure. An attacker without the correct key cannot encrypt or decrypt the message.

\textbf{Assumptions on Entities and Channels.} In the fast registration procedure, the entities related to our research include UE and AMF. We assume that the UE and AMF jointly maintain a valid security context. The communications between UE and AMF use a wireless channel. An attacker can eavesdrop the communications on the channel, but cannot decrypt the encrypted messages. In addition, the attacker can create malicious entities to send plaintext messages to UE or AMF. Only the attacker with the correct key can send the encrypted messages to UE or AMF. The UE consists of USIM and ME as mentioned above. We assume that the USIM's PIN is set to default value. Our survey showed that 83.67\% of users would not change the default PIN. And the attacker has physically access to the USIM and ME. In real life, there are indeed chances for other people to physically access the UE. For example, when a confidential meeting is held, the relevant personnel are required to put their mobile phones together outside the meeting room.

\textbf{Assumptions on Data Protections.} We assume the following data are public and available to the attacker: the default PIN and SUPI. If the operator strictly follows the 3GPP standard, the attacker cannot obtain SUPI. However, the three operators involved in this paper disable SUPI protection, which will lead to the leakage of SUPI \cite{nie2022measuring}.

\textbf{Assumptions on Authentications.} We assume that the UE and the AMF complete the mutual authentication through the security context. If the AMF can correctly process the UE's encrypted and integrity-protected registration request based on the security context, the AMF will skip the AKA procedure.
\subsection{Security Goals}
We now describe the security goals of the fast registration procedure based on the security context.

\textbf{Authentication Properties.} The 3GPP standard describes the 5G subscriber authentication properties in the document. We have identified the relevant statements and translated them into formal security goals. We use Lowe's taxonomy of authentication properties to precisely formalize the goals \cite{lowe1997hierarchy}. Lowe's taxonomy specifies four levels of authentication between two agents A and B from A’s point of view: (i) aliveness: it ensures that B has run the protocol previously, but not necessarily with A; (ii) weak agreement: it ensures that B and A have run the protocol previously, but not necessarily with the same data. This prevents impersonation attacks; (iii) non-injective agreement: it ensures that B and A have run the protocol previously and agree on the data. This prevents message tampering attacks; (iv) injective agreement: on the basis of non-injective agreement, it ensures that for each run of the protocol of A there is a unique matching run of B. This prevents replay attacks. The AMF and the UE should be able to complete mutual authentication. During the fast registration procedure, these authentications are ensured by both parties being able to complete key confirmation. And the uplink and downlink NAS COUNT values prevent replay attacks as described above. Formally, the AMF must obtain injective agreement on the fast registration procedure with the UE. Conversely, so does the UE.

\textbf{Confidentiality Properties.} While it is not clearly specified, obviously the confidentiality of the security context should be ensured. The 3GPP proposes protection measures for the security context stored in the USIM card or the baseband chip \cite{3gpp.31.102, 3gpp.33.501}. Since all the data in the security context except $K_{AMF}$ will be transmitted in plaintext, $K_{AMF}$ should be guaranteed to be confident. Formally, the cryptographic key $K_{AMF}$ protected by these measures should remain secret in the presence of the attacker.
\subsection{Overview of ProVerif}
To analyze the fast registration procedure based on the security context, we used the ProVerif prover \cite{blanchet2016modeling}. ProVerif is a protocol verification tool for the symbolic model, which is able to prove the security properties of protocols, including confidentiality properties, authentication properties and privacy properties. It has been used in verifying real-world protocols, such as 5G AKA \cite{basin2018formal}, TLS \cite{beurdouche2017messy}, FIDO UAF \cite{feng2021formal}, etc. Although ProVerif does not support XOR operation, which is important for 5G key derivation, it can still be used to verify the fast registration procedure. Since we assume that the cryptographic algorithm is secure, we can simplify the key derivation procedure.

ProVerif can automatically deduce the logical derivation of the security goals to be proved based on the formal description of the protocol. If a goal is violated, it will give the detailed attack method. In ProVerif, messages are described as terms. A term is constructed by constructors. Taking symmetric encryption as an example, we define the type key, and \textit{senc(m, k)} represents the message m encrypted using the \textit{k}, where \textit{senc(bitstring, key)} is a constructor. And the equation \textit{sdec(senc(m, k), k) = m} represents that the decryption of the plaintext with the same \textit{k}.
\begin{lstlisting}[frame=shadowbox]
type key.
fun senc(bitstring, key): bitstring.
reduc forall m: bitstring, k: key; 
    sdec(senc(m, k), k) = m.
\end{lstlisting}
\subsection{Formalizing Security Goals}
ProVerif can prove correspondence assertions, reachability properties.

Authentication properties can be checked via the correspondence assertions. The correspondence assertions are used to capture relationships between events. It can be expressed that if a specified event with some arguments has been executed, then the other with the same arguments event has been previously executed. For example, If the AMF is able to verify an encrypted and integrity-protected registration request message with information elements (including 5G-GUTI, ngKSI, uplink NAS COUNT), NAS container, and message authentication code (MAC) (event amf\_verify), then it means that the UE has initiated a unique registration request with the same parameters (event ue\_init). We can use the following query to check the injective agreement on these data.
\begin{lstlisting}[frame=shadowbox]
query x: ies, y: container, z: mac;
inj-event(amf_verify(x, y, z)) ==> inj-event(ue_init(x, y, z)).
\end{lstlisting}

Confidentiality is a reachability property. Verifying reachability properties is the most basic capability of ProVerif. The tool can prove which terms are available to an attacker by checking all possible protocol executions and attacker behaviors. Using the following query statements, ProVerif can test secrecy of the term \textit{Key} in the model.
\begin{lstlisting}[frame=shadowbox]
query attacker(Key).
\end{lstlisting}
\subsection{ProVerif Model of the Fast Registration Procedure}
We now model the fast registration procedure in ProVerif, which takes 240 lines of ProVerif code. Due to a large amount of modeling code, we only describe some key codes in this section.

\textbf{Channel and cryptographic algorithm.} We have defined a channel between UE and AMF. The channel is public under the attacks of malicious users. In addition, we also define the auxiliary cryptographic algorithms, including symmetric encryption, integrity protection, and key derivation. Specifically, the symmetric encryption and the integrity protection are used to protect the registration request message; the key derivation is used to obtain encryption and integrity protection keys from $K_{AMF}$.
\begin{lstlisting}[frame=shadowbox]
free c_ue_amf: channel.
(* symmetric encryption *)
fun senc_msg(bitstring, key): bitstring.
reduc forall x: bitstring, y: key;
    sdec_msg(senc_msg(x ,y), y) = x.
(* integrity protection *)
fun macBuilder(ies, container, key): mac.
(* key derivation *)
fun KDF_KNASenc(key): key.
fun KDF_KNASint(key): key.
\end{lstlisting}

\textbf{UE.} There are two storage methods for the security context in the UE. To do this, we pass a UEType parameter to the UE process. If the value of this parameter is ME, the security context is stored in the baseband chip. Otherwise, the security context is stored in the USIM card. In addition, to indicate that the correct PIN code is required to read the security context stored in the USIM card, we solve this problem with symmetric encryption. Specifically, the security context is encrypted with a PIN code in a symmetric encryption manner, and the correct PIN code needs to be used as a key for decryption. Similarly, for the scenario where the security context is stored in the baseband chip, we have done the same trick, but the key has become SUPI. 
\begin{lstlisting}[frame=shadowbox]
(* encrypt context with SUPI*)
fun supiContext(supi, context): context_encrypted.
reduc forall x: supi, y: context;
    getContextBySupi(supiContext(x, y), x) = y.
(* encrypt context with PIN *)
fun pinContext(pin, context): context_encrypted.
reduc forall x: pin,y: context;
    getContextByPin(pinContext(x, y), x) = y.
(* ue process *)
let ue(UEType: uetype, GUTI: guti, SUPI: supi, PIN: pin, SUPIContext: context_encrypted, PINContext: context_encrypted) = 
    if UEType = ME then (...) else (...).
\end{lstlisting}

\textbf{AMF.} We define a table to represent the correspondence between GUTI, ngKSI and security context. The AMF queries the corresponding security context according to the GUTI and ngKSI sent by the UE through the channel c\_ue\_amf. Then the AMF judges whether the registration request message is correctly encrypted and integrity protected according to the security context, and verifies whether it is replayed through the uplink NAS COUNT value. If the above checks are successful, then the event amf\_verify will be activated.
\begin{lstlisting}[frame=shadowbox]
(* table define *)
table gutiKsiContextTable(guti, ksi, context).
(* amf process *)
let amf = 
    in(c_ue_amf, m: msg);
    let InitialMessage(ClearTextIEs, NAScontainer, MAC) = m in 
    let cleartextIes(GUTI, ngKSI, xNAScount) = ClearTextIEs in
    get gutiKsiContextTable(=GUTI, =ngKSI, NASSecurityContext) in
    ...
    let xMAC: mac = macBuilder(ClearTextIEs, NAScontainer, KNASint) in 
    if xMAC = MAC then (
        let xClearTextIes: ies = sdec(NAScontainer, KNASenc) in 
        if xClearTextIes = ClearTextIEs then
        if xNASUplinkCount > NASUplinkCount then 
        event amf_verify(ClearTextIEs, NAScontainer, MAC)
    ).
\end{lstlisting}
%------------------------------------------------------------------------------
\section{Vulnerabilities}
\label{sect:vulnerability}
\begin{figure}[htb!]
	\begin{centering}
	\includegraphics[width=0.9\textwidth]{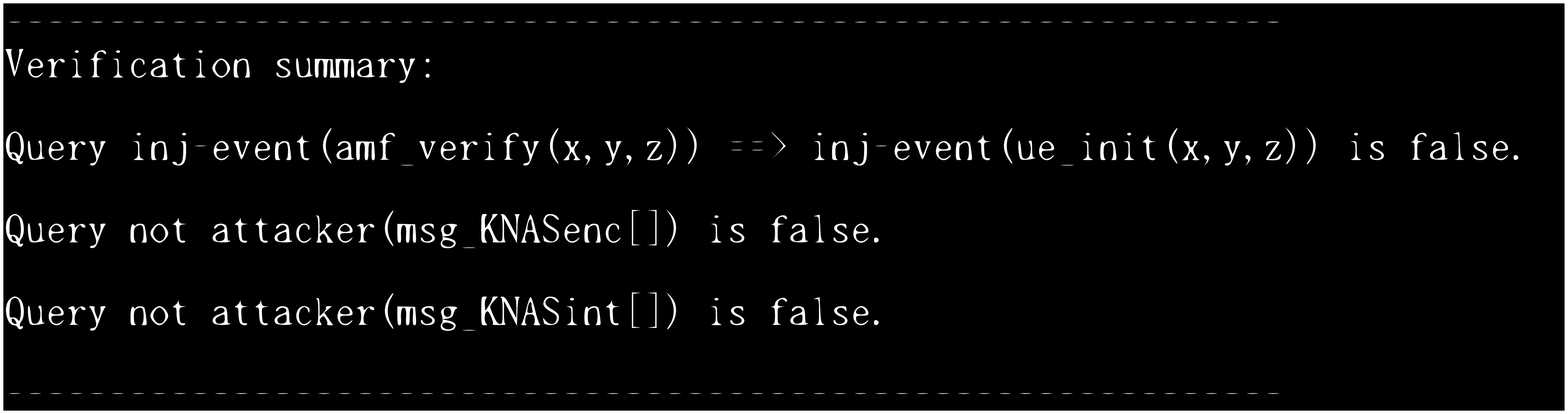}
	\caption{The formal verification results of the fast registration procedure}
	\label{fig:res}
	\end{centering}
\end{figure}

In this section, we present the formal verification results of the fast registration procedure using the security context as shown in Figure~\ref{fig:res}. It can be seen that the existing security mechanism cannot meet the security goals of the fast registration procedure. Specifically, the AMF and the UE cannot satisfy injective agreement on the fast registration procedure. That is to say, the completion of the authentication of the UE by the AMF does not mean that the UE initiates an authentication request. This shows that the attacker can impersonate the UE to complete the mutual authentication with the AMF. Moreover, an attacker can obtain keys, which defeats the confidentiality properties of the fast registration procedure. Combined with the attack paths automatically generated by ProVerif, we analyze the discovered vulnerabilities from the perspective of how the security context is stored. In addition, we tested 3 operators and found that they have vulnerabilities in both cases where the security context is stored in the USIM card or the baseband chip. Exploiting these vulnerabilities, we designed 3 attacks, which are validated on these operators. Table~\ref{findings} summarizes our findings and the results of testing on 3 operators. We denote the three operators as OP-I, OP-II, and OP-III for the privacy concern.
\begin{table}[htbp]
\caption{Summary of our findings on the security context vulnerabilities and attacks}
\begin{center}
\begin{tabular}{|c|c|c|c|c|c|}
\hline
\multirow{2}{*}{\textbf{Operator}}&\multicolumn{2}{|c|}{\textbf{Vulnerability}}&\multicolumn{3}{|c|}{\textbf{Attack}} \\
\cline{2-6} 
& \textbf{\textit{in USIM}}& \textbf{\textit{in baseband}}& \textbf{\textit{Impersonation}}&\textbf{\textit{Authentication bypass}}&\textbf{\textit{Location spoofing}} \\
\hline
OP-I & Yes & Yes & Yes & Yes & Yes  \\
\hline
OP-II & No & Yes & Yes & Yes & Yes  \\
\hline
OP-III & Yes & Yes & Yes & Yes & Yes  \\
\hline
\end{tabular}
\label{findings}
\end{center}
\end{table}

\textbf{The security context stored in the USIM card.} We analyze the operators' insecure practices regarding the security context stored in the USIM card. The 3GPP standard stipulates that the PIN-based access control can prevent the security context stored in the USIM card from being maliciously read \cite{3gpp.31.102}. However, the PIN-based access control cannot achieve the desired effect \cite{zhao2021securesim}. First, the USIM card cannot differentiate between various entities accessing the card through the PIN verification. To make matters worse, operators usually set a default PIN and disenable the PIN verification. Our survey results show that most subscribers also do not change the default PIN. Therefore, an attacker can access the security context saved in the USIM card, which uses the default PIN, with a card reader, malware, etc. In this paper, we have tested USIM cards from three operators and found that they all use 1234 as the default PIN. In addition, only one of the three operators has enhanced the USIM card to prevent the use of the card reader to obtain the security context. Mobile users of the remaining two operators may face various attacks, including traffic eavesdropping, MITM, and impersonation \cite{zhao2021securesim}.

\textbf{The security context stored in the baseband chip.} We analyze the security of the security context stored in the baseband chip from the protocol perspective. The 3GPP standard stipulates that the security context stored in the baseband chip shall correspond to the subscriber's permanent identity. It also specifies that the security context stored in the baseband chip should be deleted from the baseband chip in the following cases \cite{3gpp.33.401, 3gpp.33.501}: (1) The USIM card is removed from the ME when the ME is in power on state; (2) The ME is turned on and finds that there is no USIM card in the ME; (3) The ME is turned on and discovers that the USIM card is different from the one which was used to create the security context. However, we find that if the ME is in airplane mode or turned off, the baseband chip cannot capture the USIM card being removed or replaced. The security context stored in the baseband chip can only be used if the permanent identity from the USIM matches the identity associated with the security context. Therefore, an attacker may construct a USIM card with the same permanent identity, and replace the legitimate USIM card with the fake USIM card when the phone is in airplane mode or turned off to illegally use the stored security context.
%------------------------------------------------------------------------------
\section{Attacks}
\label{sect:attack}
In this section, we describe the proposed attacks that exploit the discovered vulnerabilities. First, we introduce our attack model and experimental setup. Then we show the two impersonation attacks by exploiting the vulnerabilities of the security context. Finally, we present two other attacks (i.e., one-tap authentication bypass and location spoofing) that result from a successful impersonation attack. We have validated them on three operators as shown in Table~\ref{findings}.
\subsection{Attack Model}
\label{sect:attackmodel}
We assume an attacker can perform attacks in the following two scenarios. In the first scenario, we assume the attacker can obtain the valid security context stored in the victim's USIM card. This can be achieved with one-time physical access through hardware (e.g., the card reader) or malware installed on the victim’s phone \cite{zhao2021securesim}. There are many types of malware, such as worms, backdoors, viruses, Trojans, that easily threaten smartphones \cite{qamar2019mobile} which can be employed by attackers to steal the private information in potential. In addition, attackers can leverage the USIM sticker to extract sensitive files from the victim’s USIM \cite{zhao2021securesim}. In this paper, we take advantage of the card reader to obtain these target parameters. In the second scenario, we assume the attacker can obtain a mobile device, which stores a valid security context of the victim. This does work in real life. For example, on some special occasions, several people will take turns using a mobile phone, and each person will use a different USIM card. An attacker who is one of the members can then attack the previous user. Alternatively, the attacker can trick the victim into inserting their USIM card into the attacker's phone in front of the victim and then return the USIM card. The attacker only needs to ensure that the phone is turned off or in airplane mode when the victim's USIM card is pulled out. It is worth noting that attackers can use their own USIM cards to launch the above attack and forge their location information as described in Section~\ref{locspoof}. Moreover, we assume that the attacker needs to know some basic knowledge about 4G/5G, and have the ability to use the card reader to read and write the USIM card.

\textbf{Ethics discussion.} We perform all the experiments in a responsible way: all USIM cards and mobile devices involved in the experiments are our own to prevent affecting other users. We have notified operators about the unsafe implementations. At the same time, we have submitted the discovered protocol vulnerability to the GSMA for mitigation and obtained a coordinated vulnerability disclosure (CVD) number CVD-2022-0057.
\begin{figure}[htb!]
	\begin{centering}
	\includegraphics[width=0.5\textwidth]{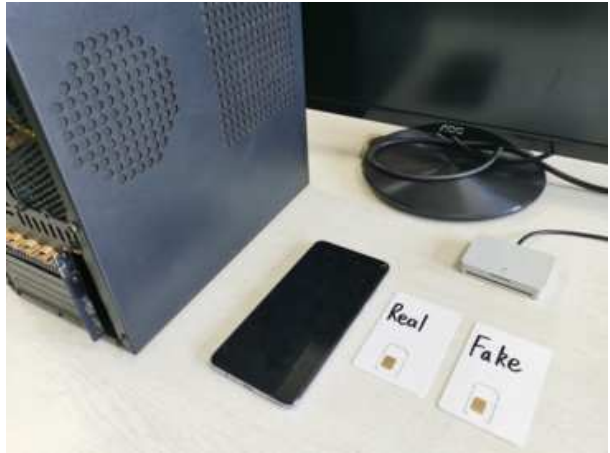}
	\caption{Our experimental setup}
	\label{fig:setup}
	\end{centering}
\end{figure}
\subsection{Experimental Setup}
As Figure~\ref{fig:setup} shows, our setup consists of a computer with the Windows operating system. The computer is equipped with the software which sends APDUs to interact with the USIM card, and cooperates with a card reader to read and write the USIM card. In addition, we used a Huawei Mate30 5G as the test phone, which can connect to the 4G and 5G networks of the three operators. There are two kinds of USIM cards used in our experiments. One is a legal 5G USIM card that is assigned by the operator and facilitates the subscriber to access the 4G/5G network (we name it a real card). And the other is a programmable USIM card (we named it a fake card). The files (e.g., IMSI, EPSLOCI and EPSNSC) stored in the fake card can be read and written at will.

We found that when using the real card to access the operator's 5G network, the security context is stored in the baseband chip, and when accessing the 4G network the security context is stored in the USIM card. Therefore, in order to launch an impersonation attack using the security context in USIM, we choose to access the 4G network. And when verifying the impersonation attack using the security context in the baseband chip, we choose to access the 5G network.
\subsection{Impersonation Using the Security Context in USIM}
The PIN-based access control cannot effectively protect the security context stored in the USIM card. If the operator does not enhance the security of the USIM, an attacker can use hardware or software to obtain these security contexts. Then the attacker can leverage the vulnerability and the fast registration procedure to perform several attacks, such as traffic eavesdropping, MITM attack, and impersonation \cite{zhao2021securesim}. The impersonation attack described in \cite{zhao2021securesim} is implemented using the open-source software srsLTE \cite{gomez2016srslte}. However, since the software utilizes software-defined radio (SDR) to send wireless signals, it is not as stable as mobile devices equipped with a baseband chip. In this paper, our impersonation attack based on this vulnerability is performed with a commercial mobile device.

\begin{figure}[htb!]
	\begin{centering}
	\includegraphics[width=0.6\textwidth]{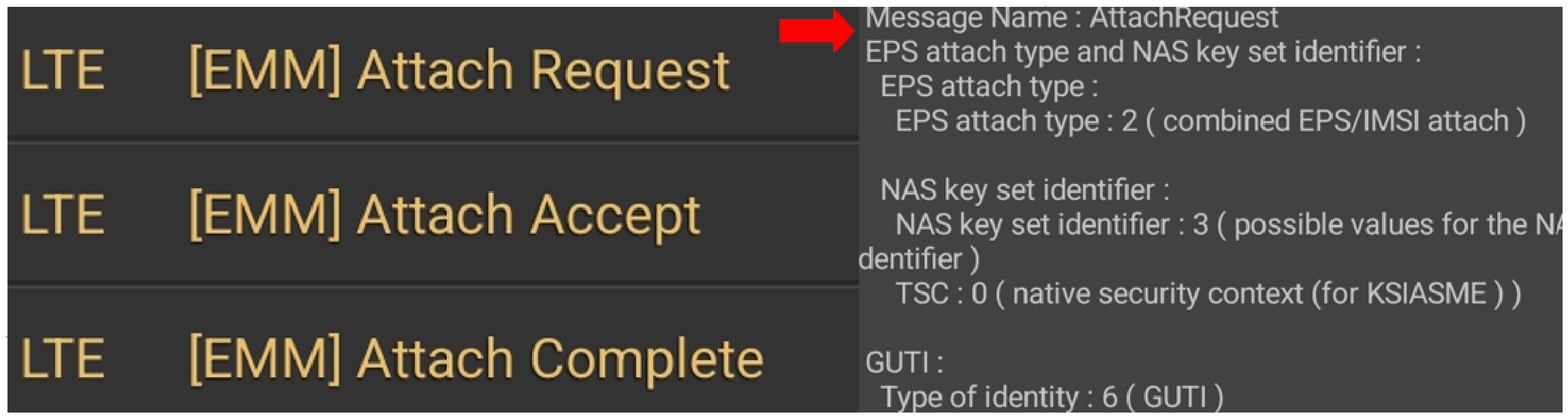}
	\caption{The NAS signaling messages of the impersonation attack in a 4G network}
	\label{fig:4G}
	\end{centering}
\end{figure}

\textbf{Attack Procedure.} We assume that the victim has connected to the operator's 4G network using a mobile phone equipped with the real card. At this point, a valid 4G security context is already stored in the real card. The attacker can then launch the impersonation attack. First, the attacker uses the card reader to obtain the files (i.e., IMSI, EPSLOCI and EPSNSC) required for the fast registration procedure. Second, the attacker uses the same method to write these files into the fake card. Then the attacker inserts the fake card into the phone and uses the fast registration service to access the operator’s 4G network as the victim. Figure~\ref{fig:4G} shows the NAS signaling messages. The attach request signaling includes the NAS key set identifier (in file EPSNSC) and the GUTI (in file EPSLOCI), and the 4G network does not initiate an AKA procedure. Both sides use the key in file EPSNSC to derive various keys to ensure the security of subsequent communications.

\textbf{Discussion.} We find the attack poses real threats to subscribers of OP-I and OP-III. OP-II has made security enhancements to the USIM card to prevent malicious reading of the security context. We set the IMSI of the fake card as the victim's permanent identity in order to be able to properly handle the identity requests that the network sometimes initiates. Additionally, we find that the NAS security context persists for a long time until a new AKA process occurs. But each registration procedure will change the GUTI. To launch the attack for a long time, the attacker needs to sniff the signaling to obtain the latest GUTI of the victim \cite{hong2018guti}. Although we launch the attack under the 4G network when the security context is stored in the USIM card, the experimental results from the 4G network are still reasonable in the 5G network. The reasons are presented as follows: (1) The 4G USIM card and the 5G USIM card have the same file system, which sets different access conditions for different files to ensure security \cite{3gpp.31.102}. The 5G USIM card stores the 5G security context in the form of a file, and its access condition is set to PIN, which is the same as the access condition of the 4G security context in the 4G USIM card. Therefore, an attacker can use the card reader to obtain the 5G security context stored in the 5G USIM card. (2) The fast registration procedure based on the security context in 4G and 5G networks have the same process \cite{3gpp.24.501, 3gpp.24.301}.
\subsection{Impersonation Using the Security Context in Baseband Chip}
The security protection proposed by the 3GPP standard for the security context stored in the baseband chip cannot achieve its intended purpose. First, when the mobile phone is in airplane mode or turned off, the baseband chip cannot determine whether the mobile phone card has been pulled out or replaced. Secondly, the baseband chip decides whether to use the security context for fast registration procedure by judging whether the permanent identity in the USIM card is consistent with the identity corresponding to the saved security context. In this paper, we propose an impersonation attack that exploits the security context stored in the baseband chip.

\begin{figure}[htb!]
	\begin{centering}
	\includegraphics[width=0.6\textwidth]{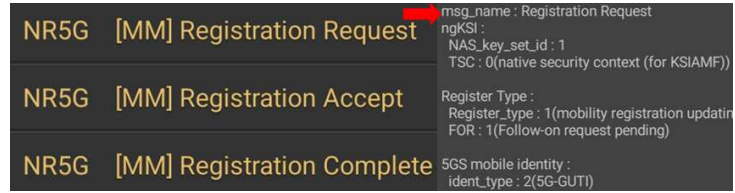}
	\caption{The NAS signaling messages of the impersonation attack in a 5G network}
	\label{fig:5G}
	\end{centering}
\end{figure}

\textbf{Attack Procedure.} First, the attacker determines the permanent identity SUPI of the victim's USIM card. Since these three operators have not enabled protection against SUPI, SUPI is transmitted over the air interface in the form of IMSI at this time \cite{nie2022measuring, lilly2017imsi}. Then the attacker writes it to the fake card. Second, the attacker would need to obtain a phone with a legitimate security context. It is indeed possible as described in Section~\ref{sect:attackmodel}. Third, the attacker inserts the fake card into the phone, and then turns off airplane mode or turns the phone on. Since the permanent identity in the fake card is consistent with the identity corresponding to the security context, the mobile phone determines that the saved security context is valid, and uses it to initiate a fast registration procedure. Finally, the attacker can use the fake card to access the operator's 5G network as shown in Figure~\ref{fig:5G}. 

\textbf{Discussion.} Since this impersonation attack is done by exploiting a protocol vulnerability, it is applicable to the three operators that provide fast registration services. The attacker needs to ensure that the victim is no longer connected to the network during the attack. Otherwise, the security context saved in the phone will be invalid. However, attackers can launch the One-tap authentication bypass attack discussed below within seconds of accessing the network. Thereafter, although the victim will reconnect to the network, which ends the impersonation attack, the attacker was already logged into the victim's application account at this point.
\subsection{Other Attacks}
\label{locspoof}
After the attacker has successfully launched the impersonation attack, the attacker can also launch some other attacks, including one-tap authentication bypass and location spoofing.

\textbf{One-tap authentication bypass.} The one-tap authentication service allows users to quickly sign up or log in to their application accounts using the tokens provided by the operator \cite{zhou2022simulation}. All three operators discussed in this paper support this service. The service is based on cellular network status. The attacker's device needs to have the same cellular network state as the victim to bypass this authentication. Since the attacker can launch impersonation attacks, the attacker can bypass this authentication to log into the victim's application account. This poses a great threat to user privacy.

\textbf{Location spoofing.} Operators can locate subscribers based on which base station the subscribers are connected to. However, the attacker with the security context can launch impersonation attacks far from the victim, which will challenge operators to locate users. Moreover, this could also be exploited by criminals to forge evidence.
%------------------------------------------------------------------------------
\section{Survey}
\label{sect:evaluation}
\begin{figure}[htb!]
	\begin{centering}
	\includegraphics[width=\textwidth]{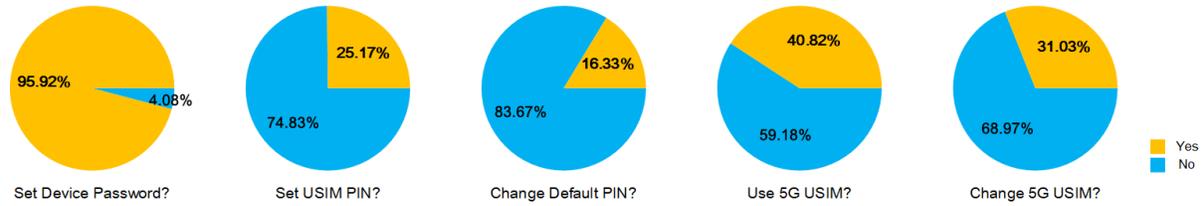}
	\caption{The USIM usage survey}
	\label{fig:survey}
	\end{centering}
\end{figure}

To evaluate the range of users that the impersonation attacks discussed in this paper can affect, we have conducted an informal survey and obtained responses from 147 volunteers. Figure~\ref{fig:survey} shows the results of this survey. The results show that impersonation attacks are feasible in practice. Specifically, 95.92\% of the participants set the password for their devices, but only 25.17\% of the users enable the PIN verification for their USIM cards. To make matters worse, among the users who enable the PIN verification, 83.67\% of the subscribers use the default PIN. The security context stored in the USIM of these users may be exposed to attackers. In addition, 59.18\% of subscribers are still using 4G USIM cards, and 68.97\% of them are unwilling to upgrade to 5G USIM cards. However, the 5G USIM cards of the 3 operators currently do not support storing 5G security contexts. Therefore, when using the USIM cards (including 4G USIM and 5G USIM) of these 3 operators to access the 5G network, the 5G security context is stored in the mobile phone. This may be exploited by attackers to launch impersonation attacks.
%------------------------------------------------------------------------------
\section{Countermeasures}
\label{sect:countermeasure}
The vulnerabilities stem from the design flaw of the 3GPP protocol and the unsafe practices by operators. We propose the following countermeasures from four perspectives.

\textbf{The operator.} First, the operator should enhance the security of the USIM card to prevent the security context saved in the USIM card from being illegally read. Second, operators can disable the fast registration services. For each registration request, the network initiates an AKA procedure to ensure security. Moreover, for the UE that has already accessed the network, it is better for the operator to reinitiate an AKA procedure at regular intervals to update the security context. Third, the 5G USIM card should be enhanced to support storing the 5G security context.

\textbf{The user.} On the one hand, users can enable PIN verification and modify the default PIN code at the same time. On the other hand, after the operator enhances the 5G USIM card, it is better for users to upgrade to the latest 5G USIM. Not only does this prevent the security context from being stored in the baseband chip, but it also leverages the security features of 5G to protect its permanent identity \cite{3gpp.33.501}.

\textbf{The 3GPP protocol.} Checking the validity of the security context stored in the baseband chip should incorporate other parameters (e.g., ICCID), not just the permanent identity. This will greatly increase the difficulty of the attack.

\textbf{The equipment manufacturer.} Mobile phone manufacturers and baseband chip manufacturers should coordinate the transmission of USIM card status information. When the mobile phone is in airplane mode or turned off, the baseband chip should be able to know the change of the USIM card status information, and then decide whether to delete the security context.
%------------------------------------------------------------------------------
\section{Related Work}
\label{sect:relatework}
We discuss the related work with the following two categories, namely, impersonation attacks in mobile networks, and security research on the USIM card.

\textbf{Impersonation attacks in mobile networks.} Meyer et al. \cite{meyer2004impact} present that an attacker could exploit the inherent authentication flaws of the GSM network to imitate the victim's access to the GSM network. Rupprecht et al. \cite{rupprecht2020imp4gt} propose the impersonation attack that exploits the lack of integrity protection on the user plane and the reflection mechanism of Internet Control Message Protocol (ICMP). Zheng et al. \cite{zheng2017ghost} find one vulnerability in Circuit Switched Fallback (CSFB) where the authentication step is missing, which allows an attacker to impersonate a victim. Cui et al. \cite{cui2022security} point out that the attacker can impersonate victims to register to IP Multimedia Subsystem (IMS) by exploiting the 3GPP protocol flaws and unsecure practices of operators. Most relevant to our research is the impersonation attack based on the security context stored in the USIM card \cite{zhao2021securesim}. Unlike them, the impersonation attack presented in this paper is implemented using real mobile phones, not SDR. The impersonation attack using mobile phones has the following advantages: (i) a more stable signal; (ii) the ability to further launch other attacks, such as One-tap authentication bypass.

\textbf{Security research on the USIM card.} Liu et al. \cite{liu2015small} present how to copy a 3G/4G USIM card via the side-channel attack. An attacker can launch the SIMjacker attack to track user location without the user's awareness \cite{simjacker}. Chitroub et al. \cite{chitroub2018sim} study the security mechanisms of the embedded USIM (eSIM) and arise the potential vulnerabilities, which can be exploited to exhaust the memory of eSIM. Zhao et al. \cite{zhao2021securesim} uncover three vulnerabilities of the PIN-based access control of the USIM card. In this paper, we analyze the security context generated by the USIM card.
%------------------------------------------------------------------------------
\section{Conclusion}
\label{sect:conclusion}
The security context is critical to mobile users, since it allows a mobile user to access the 5G network through a fast registration procedure. However, the previous research on the security context only considers the case where it is stored in the USIM card, but ignores the fact that the security context can be also stored in the baseband chip. And these studies were conducted through manual analysis. In this work, we consider both of the above scenarios, and use the formal approach to model a fast registration procedure based on the security context. We find two vulnerabilities, which are caused by the insecure implementations of operators or the 3GPP protocol flaws. We propose two novel impersonation attacks based on these vulnerabilities and confirm them in three operators. Furthermore, we also discuss two other security threats that can arise from impersonation attacks. Finally, we propose several countermeasures from multiple perspectives to defend against the attack.
%------------------------------------------------------------------------------
\section* {Acknowledgments}
This work is supported by the National Natural Science Foundation of China (No. 62001055 and 61872386), and the Beijing University of Posts and Telecommunications-China Mobile Research Institute Joint Innovation Center.
%------------------------------------------------------------------------------
%
\label{sect:bib}
\bibliographystyle{unsrt}
\bibliography{ref-mobisec22}
%------------------------------------------------------------------------------
\end{document}